# Scaling Culture in Blockchain Gaming:
## Generative AI and Pseudonymous Engagement

PRE-PRINT

Henrik Axelsen, University of Copenhagen, Denmark.
E-mail: heax@di.ku.dk.

Sebastian Axelsen, Independent
E-mail: s.axelsen89@gmail.com

Valdemar Licht, University of Copenhagen, Denmark.
E-mail: valdemarlicht@hotmail.com

Jason Potts, RMIT University Melbourne, Australia
E-mail: jason.potts@rmit.edu.au

Abstract:

Managing rapidly growing decentralized gaming communities brings unique challenges at the nexus of cultural economics and technology. This paper introduces a streamlined analytical framework that utilizes Large Language Models (LLMs), in this instance open-access generative pre-trained transformer (GPT) models, offering an efficient solution with deeper insights into community dynamics. The framework aids moderators in identifying pseudonymous actor intent, moderating toxic behavior, rewarding desired actions to avoid unintended consequences of blockchain-based gaming, and gauging community sentiment as communities venture into metaverse platforms and plan for hypergrowth. This framework strengthens community controls, eases onboarding, and promotes a common moral mission across communities while reducing agency costs by 95 pct. Highlighting the transformative role of generative AI, the paper emphasizes its potential to redefine the cost of cultural production. It showcases the utility of GPTs in digital community management, expanding their implications in cultural economics and transmedia storytelling.

Keywords

# Introduction

In the intersection of cultural economics and technological innovation, the management and understanding of rapidly scaling cultural groups presents novel challenges. These challenges include (i) maintaining cultural identity amid hypergrowth, (ii) managing pseudonymous interactions that can obscure intent and behavior, (iii) ensuring equitable value distribution within decentralized structures, (iv) adapting to the evolving norms and ethics of a growing community, and (v) mitigating risks associated with the integration of blockchain technologies and their economic implications.

The economic problem we focus on in this paper is the scaling of the production of culture during hypergrowth (Izosimov, 2008) using the example of blockchain based gaming. This is hard, especially when the core production is culture, and the business expands rapidly; the annual growth rate hits high double-digits, and human capital needs grow dramatically (Valencia, 2019).

The cost function of human community curation encapsulates the monetary expenses tied to salaries and tools, the time spent on onboarding and content review, and the emotional toll from managing negativity and making continuous decisions, coupled with indirect costs related to reputation and potential legal risks. Organizations need the appropriate personnel and structures to drive their rapid growth. These are necessary to stay healthy. This is expensive, and balancing these costs often necessitates a blend of human and automated interventions. This is also the case for virtual blockchain-based gaming startups and generally any startup, including games organized in Decentralized Autonomous Organizations (DAOs), such as Decentraland.

This paper is about the innovation of using AI to perform that function better. We explore this problem with blockchain-based gaming communities adopting a device-agnostic metaverse model and planning for hypergrowth across different platforms, blockchains, and brands. We introduce a novel method of cultural analytics leveraging a class of Large Language Models (LLM), specifically a suite of open-access Generative Pre-Trained Transformer (GPT) models to address the following research question: "Are LLMs an option for identifying, nurturing, and sustaining culture in gaming communities going through hypergrowth?"

We considered four hypotheses: (i) pseudonymous actor intent can be predicted from online activity, (ii) large language models can make mass content moderation effective and efficient, (iii) a common moral mission can be implemented across communities, and (iv) there is a huge automation potential to reduce moderation costs in off-chain governance of decentralized communities.

Our findings suggest that the combination of AI and Blockchain tech stacks hold significant potential to solve the problem of scaling culture production in blockchain-based business models while reducing agency costs. By extrapolating our findings, we offer the following general propositions on the impact of using open-access generative AI to support cultural production in growth:

(i) Cost-effective open-access Large Language Models (LLMs), such as OpenAI's GPT, present an opportunity to automate innovation in managing societal and cultural groups. (ii) LLMs highlight a spectrum of decentralization in the blockchain and internet-based creator economy, with tools like GPT offering both fully and semi-decentralized strategies for curating experiences. (iii) Their integration with blockchain gaming can streamline moderation and cultural preservation, adaptable across games. However, (iv) generative AI's capacity to craft narratives could challenge human uniqueness and face potential backlash. Moreover, (v) businesses' reliance on centralized APIs, especially in a decentralized context, introduces governance and operational risks and potential privacy concerns, urging caution in long-term commitments.

We contribute to the transmedia storytelling discourse with (i) the application of a contribution-based business model focused on collaboration, teamwork, and contribution around a common moral mission when transforming from web2 to a decentralized, blockchain-based internet, the so-called web3, (ii) the governance of blockchain-based business models by introducing generative AI-based supported community management policies, (iii) meta-sustainability by internalizing sustainability values at an individual level while developing a collective mindset that understands the importance of doing the right things the right.

## Method

*Background to case study*

Blockchain-based gaming promises to counterbalance power in production, enabling players to earn and retain value, and offer enhanced openness and integration between games and their underlying blockchains (Egliston & Carter, 2023), and major brands increasingly want to participate in the emerging disruptive business models (X. Nguyen, 2023). Such crypto gaming use tokenization and smart contracts and are reshaping the gaming economy by allowing players to monetize assets and gain enhanced control over their experiences.

Cultural science suggests that culture, through its evolutionary and complex nature, facilitates group formation and stimulates knowledge and innovation and promotes social cohesion, where shared practices and symbols within groups reinforce identity (Hartley & Potts, 2014). Platforms like Discord are essential in fostering this digital interconnectedness in decentralized models by transforming players from consumers in traditional pay-to-play (P2P) games into producers and owners of blockchain-issued non-fungible tokens (NFTs). As players profit through play-to-earn (P2E) mechanisms, akin to stock options in traditional firms, they also have the potential for real-world trade. However, the dominance of such profit-driven incentives may pose risks (Delic & Delfabbro, 2022), including potential exploitation and mental health concerns, such as technostress from addiction, urging developers to monitor and address potential harm.

With the emergence of web3, with no centralized authority controlling or regulating the network, Decentralized Autonomous Organizations (DAOs) (Santana & Albareda, 2022) are proliferating as an emerging organizational form to support creator economies, also in gaming (Egliston & Carter, 2023; Glaveski, 2022). DAOs operate as decentralized, fluid organizations through token-based working practices with pseudonymous actors, making it difficult to define relevant externalities (Axelsen et al., 2022). Yet, despite the DAO concept and typology being nascent and still developing, it has been suggested that DAOs could further confront the societal problems plaguing the metaverse, the current focus of many web3 gaming studios, by providing incentives to drive behavior in the required direction, with decentralized control of data, a greater level of user engagement, ownership and involvement (American Enterprise Institute, 2022).

Decentralized community projects, including DAOs, could be considered projects to create digestible stories. "The narrative must fit the community as well as the project" (Shorin et al., 2021), and the core community members must believe that community moderators manage the story, the mission, and divergent stakeholder interests. Measures must be implemented to scale, as the case of CityDAO suggests (*CityDAO*, 2023), where a tokenized collective experienced hypergrowth, after some crypto notabilities purchased NFT-citizenships in 2021, resulting in temporary sellout, an unsurmountable workload, and the DAO deciding to halt most external operations, while building infrastructure to allow better scaling (Taken, 2022).

*Participants*

In 2017, the subject of this case study, gaming studio startup Reality+, forked the Ethereum blockchain to create unique approach to brand specific NFT collections for card gaming in a so-called free-to-play (F2P) model. As they now transition to a metaverse model with customers scaling globally, their NFTs grant players avatars, seasonal awards, and other perks, and Reality+ is increasingly becoming a metaverse service for mega-brands that aim to amplify their narratives through transmedia storytelling (Perryman, 2008).

With around 250,000 current users and a projected 5 million by 2024, Reality+ seeks a strategic approach to manage its growing community and ensure sustainable scaling into what observers estimate will become a US$ 8-13 Trillion business by 2030 (Burke, 2022). Among its customers are BBC Studios (Doctor Who and recently 26 other BBC brands), ITV (Thunderbirds, Love Island).

To advance the creator economy surrounding their web3 gaming models, Reality+ considered establishing a DAO to harness its community's affinity as it continued its high-growth journey, while distributing fair profits to community contributors. Yet, while the digital work practices of the communities in the gaming platform operate very decentralized and are like those in DAOs, the DAO concept and decentralized governance conflicts with the need for the customer mega-brands to remain in control of their brand IP. Hence, Reality+ is experimenting how to combine a decentralized gaming platform with traditional business IP-controls.

The most popular game in Reality+' universe is Doctor Who Worlds Apart (DWWA), based on the world's longest-running sci-fi series (Bell, 2020) that has held significant cultural influence in the UK and beyond for the past 60 years. With numerous global fan communities, Reality+ emphasizes the Doctor Who gaming card and web3 experience. This gaming experience also serves as the blueprint for the entire Reality+ game ecosystem, attracting other brands seeking web3 and metaverse experiences.

At the time of research, a dedicated gamer spends US$1000 annual recurring revenues (ARR) in line with mature video gaming, some 27x higher than casual gamers in traditional F2P trading card gaming. The F2P model is subject to high churn in line with the industry, but also high net growth.

### *The Community and the Metaverse*

Out of approximately 250,000 wallets on the Reality+ NFT platform, about 75,000 are tied to DWWA. Notably, a subset of around 3,000 form the gaming community, with 1,100 actively participating in discussions. A team of 30 part-time moderators oversees this community. However, in addition to the risk of a "CityDAO" experience, there is a concern that unchecked growth may dilute the community's appeal, especially as associated brands like BBC expand their global reach via platforms like Disney+.

Reality+ aims for device-agnostic operations, but currently revolve around Discord for community management, with games tied to the Ethereum blockchain. As they migrate into metaverse platforms they aim to embed a worldview promoting ethical avatar behavior (Park & Kim, 2022). Yet, advanced metaverse platforms like Sandbox and Decentraland blur online and offline personas through decentralized structures, blockchain, and DAO governance (Dwivedi et al., 2022; Egliston & Carter, 2023), and despite advancements, a fully immersive metaverse remains elusive, but the direction is clear: In a metaverse context, composability and interoperability, using DeFi protocols (Werner et al., 2021), are essential.

Contrary to the general crypto industry, Reality+ is also cautious of potential regulatory issues around DeFi..

In this context, the DWWA card game epitomizes Nguyen's "art of agency" theory (C. T. Nguyen, 2019). It offers players dynamic narratives, profound moral choices, and collaborative experiences, all while maintaining game integrity and compliance with external stakeholder expectations. Drawing from DWWA, Reality+ has established a compelling gaming approach, where web3 assets and real-world connections are fortified and now enrich partnerships with sports leaders like FIFA and Velon by offering players a harmonized blend of strategy and story, operating within distinct communities, yet linked to an overall platform with a common moral mission.

*Community challenges*

Community challenges and requirements were identified through semi-structured interviews:

1. Scaling a factor 20X over the next year while avoiding the chaos CityDAO experienced and keeping curation costs low.
2. Understanding who customers are vs. stakeholders/creators and how to effectively spot these. Although segmentation of online gaming motivation using analytical methods is not new (Kahila et al., 2023; Tseng, 2011), the moderators have so far been unable to model this, failing to identify the relevant annotation variables; but having a clear view of what personas they are looking for – crypto investors and dedicated gamers. They observe valuable community and game contributions but cannot identify in advance who, when, and where the contribution is delivered and with what intent before it has been delivered. Understanding intent a priori would help moderators greatly in providing the relevant onboarding procedure.
3. Increased risk from toxicity in gaming environments entering a metaverse scaling journey encapsulating other franchises requires smarter ways to moderate community chat effectively.
4. Although most community members mirror the Doctor Who characters' behavior and agree that this is a moral compass for them, some members are more extreme, representing cult fandom, which may cause division around innovation.
5. Understanding culture production and how to create internal controls for managing stakeholder expectations.

6. Institutionalizing a balanced incentive model to address the unintended consequences of the new business model.

Specific to the scaling plans and need for more resilient infrastructure, the moderators and founders delineated the set of requirements as follows:

1. Design a model development framework to create a cost-effective rapid deployment classification system for cultural aspects of the community, using a cultural science approach to identify improvements needed for scaling/hypergrowth.
2. Develop a suite of analytical models with good to very good F1-scores to understand, automate, and flag to moderators for their further curation (a) intent of pseudonymous actors and classify stakeholders for proper onboarding, (b) unwanted behavior to penalize/moderate, (c) desired behavior to reward with NFTs or other perks for meaningful contributions as part of a "proof of contribution" concept.
3. Outline the community's sentiment and key drivers of "meaningfulness" and assess the general cultural sentiment to enable understanding and planning for a resilient growth journey.

## Procedure and Apparatus

For interviews and workshops, we chose an explorative, qualitative approach to the classification of cultural practices leveraging the cultural science method. This process led us to workshop critical elements of the cultural practices within the community and how to assess their maturity and identify opportunities for improvement. The unit of analysis was the practices undertaken by the community, the story of the Doctor Who franchise, its characters, how the show had affected the community culture, and the developing plans to migrate the game to a metaverse platform while preparing for hypergrowth.

Moderators supplied us with chat data from the community's Discord forum to investigate the potential of using AI to automate and improve community management. Since its inception 18 months prior, the data scrape include over 65,000 chats. A data consent form was shared to comply with GDPR requirements for research projects. In a subsequent replication experiment to another Reality+ game community, Thunderbirds, another long running British science fiction series, a data scrape covering 102,000 chats was provided.

For the analytics part, large language models (LLM) models easily align with user needs and human feedback (Ouyang et al., 2022) to classify a range of issues, including the assessment of intent by leveraging language processing to analyze large volumes of text, for example from online communities, using embeddings for search, clustering, recommendations, anomaly detection, diversity measurement or classification (*OpenAI API*, 2023).

Where previously, it was necessary to fine-tune LLMs on datasets of thousands of examples for good performance, OpenAI's recent launch of GPT3 and -4 allow scaling up a language model with only a few examples (few-shot learning) to achieve similar performance (Brown et al., 2020).

Creating an LLM is not a small task but requires a specialized programming team and expensive hardware with a powerful GPU to run the model and handle a big dataset. There are several ways to model the problems analytically, which could likely all yield good results but would come with different risk/return trade-offs, which were discussed with the moderators and all boil down to a question of data availability, performance, and cost:

1. Train our own natural language classifier using, e.g., word embeddings and Long Short-Term Term Memory (LSTM) architecture.
2. Use a zero-shot LLM (no training required) such as OpenAI's text-davinci model.
3. Fine-tune an existing LLM (few-shot learning) such as OpenAI's ada model.

The recent rise in model complexity and limited open-source access have spurred model acceleration methods, notably knowledge distillation. This technique has a larger model teach a smaller one, allowing cost-efficient models to harness the advantages of bigger models for tasks needing less complexity. Such practices, like in-context and few-shot learning, are gaining traction among researchers aiming for content analysis. Given the limited team size and resources, our goal was to find less resource-intensive alternatives to traditional analytical processes.

Open-access GPT models could automate community moderation aspects, making it cost-effective for new communities. Leveraging OpenAI's GPT3 API was ideal due to its model variety and potential for low-cost prototyping, deferring the bulk cost until production. The GPT-3 model suite is apt for zero-shot learning. However, the specific weights of these proprietary models remain undisclosed, and our approach therefore technically is not true distillation, but applies distillation principles with zero-shot learning, creating an initial dataset that is then curated by humans. Through iterative fine-tuning, more examples are generated and validated by moderators, mirroring traditional taxonomy development but faster and cheaper.

### Apparatus

*Rapid development framework*

We designed the following procedural steps to rapidly create a suite of models based on the API:

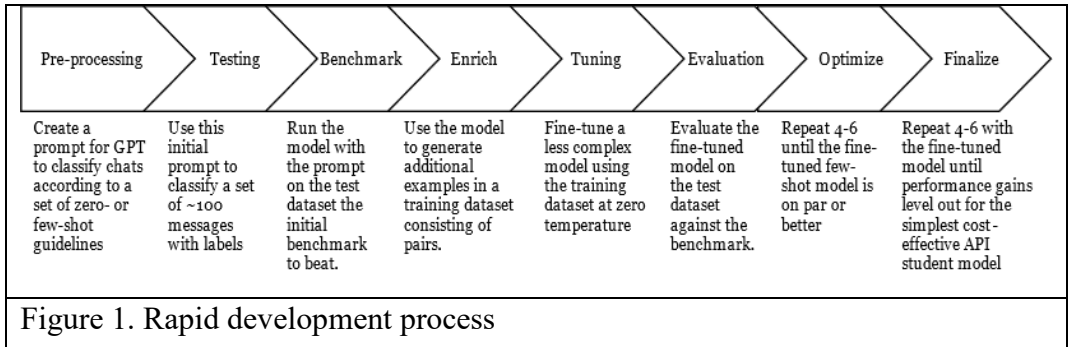

Figure 1. Rapid development process

If a model is not performing meaningfully after step 3, we would first add additional manually annotated chat examples to the training dataset and re-train the model on a larger sample of training data. If the model is still not performing, it may be that the problem is not a classification problem. In this case, first, we let more than one person classify and validate the dataset to check the consistency of human annotation. If the answers are subjective, it can be challenging to evaluate the success of each response using classification and labeling, but it may require prompt engineering. Tasks such as open-question answering may involve manual grading and curation. This can be a concern since human graders bring their biases to the task. To reduce this, multiple graders and consensus metrics were needed for the curation process, implemented through detailed annotation guidelines, validated by the moderators.

Figure 2 provides an overview of the framework architecture, linking off-chain cultural practices through models, curation, and NFTs.

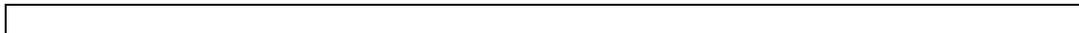

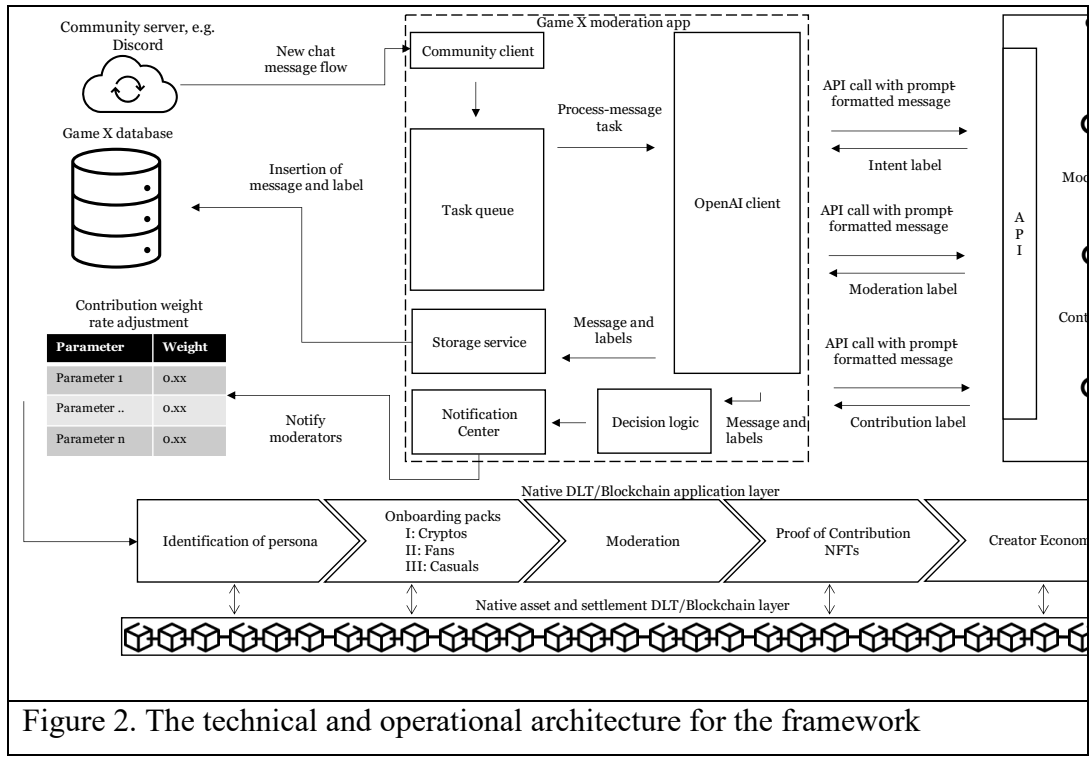

Figure 2. The technical and operational architecture for the framework

The community moderation teams had previously without success utilized analytical and cartographic tools for community management within Discord, yet the identification of key personas and contributions remained challenging. Our primary objective was to devise a predictive model to classify message intent in the Doctor Who community based on user activity and then convert this to user intent through a rules-based approach. Collaborative workshops with the founder and moderator team resulted in the identification of three primary personas in the Reality+ gaming community:

- Cryptos: (a) NFT speculators who engage in community dialogue and anticipate a return on investment, and (b) Flippers, short-term investors with minimal community contributions but potential future speculators.
- Fans: Collectors emotionally attached to their cards, adding significant value to the community. This aligns typology-wise to "strategizers" in a metagaming context (Kahila et al., 2023).
- Casuals: Players focused on gameplay rather than NFTs, with minimal community contributions.

Moderators noted consistent persona behavior across games and geographies, suggesting shared cultural tendencies, implying an overarching Reality+ ethos transcending Doctor Who's core fandom. To classify intent, we refined an OpenAI ada model using 463 manually labeled and curated chat messages and validated its performance against a test set of 116 labeled chats. We later tested this on another game, Thunderbirds, as further outlined below.

*Moderation model*

User perceptions of community moderation can influence norm enforcement in social media, making it crucial for automated moderation to be viewed as fair and just by the community (Gonçalves et al., 2021; Myers West, 2018). OpenAI's GPT series has a built-in moderation model for toxic content, offering added value. Yet, testing revealed issues like flagging gaming or blockchain terms as violent and overlooking spam leading us to reject this option. We trained the GPT model using the public Kaggle "Toxic Comment Classification Challenge" dataset to address this. Initially, attempting a model using only the Kaggle data led to overfitting. Integrating past moderation data from images of deleted chats, presented another challenge, which was overcome with the use of open-source extracted using Optical Character Recognition software. We initially trained the model using eight predefined labels from both datasets. Still, the model remained overly sensitive to certain keywords.

A 3$^{rd}$ iteration with reduced labels prioritized context over specific terms, showing improvement. We then tried a reduced set of labels, as the practical use test was flagged for further human curation anyway. The rationale was to reduce the model's dependency on the labels and look further into the context of the message as such, not the label. This model performed better than the first iteration but still did not meaningfully capture the community language. In the final model, we then replaced the not_toxic messages in the dataset with actual community chat examples from the Discord database, which significantly reduced the false-positive rate and adjusted the model's ability to comprehend how these keywords could be used in a non-toxic manner. The final dataset for testing consisted of 383 out-of-sample chats.

*Contribution model*

Historical reward instances, termed "fatfingers" in the community—referring to airdrops from the co-founders' Discord handle, the "FatViking", were initially gathered to develop the model. These rewards were highly valued by the community as these were often rare NFTs that could accrue significant financial or status value. However, most examples were private appreciations from FatViking rather than public chat messages suitable for model training, but the data provided a foundation for zero-shot modeling, highlighting valuable contributions such as:

- Reporting bots/spam links/fake accounts,

- Educating newcomers on the trading card game, blockchain, and NFTs,
- Sharing crypto and airdrop updates,
- Expertise in Doctor Who characters and lore,
- Welcoming and providing constructive advice,
- Updating the community on Doctor Who news,
- Promoting the Doctor Who trading card game,
- Offering support to new members,
- Steering conversations positively.

This led to a 2-phase development approach:

1. Test OpenAI's chatGPT3 and -4 to tag messages in the dataset exhibiting the behavior outlined.
2. Design a proactive contribution policy grounded in FatViking's past actions and enriched by behaviors vital for growth and institutionalizing the "proof of contribution" idea. The model should assign varying weights to diverse attributes, enabling moderators to guide and adjust rewarded contributions. Additionally, it should allow contributions to build over time, potentially transitioning to decentralized community reputation systems like (*SourceCred*, 2023) or (*Coordinape*, 2023) or even integrating into future algorithmic governance models through smart contracts, once the ecosystem has mature and curation becomes consistent across communities to allow for full automation without human curation.

The new contribution policy was designed based on a series of consultations. To ensure holistic coverage of the policy objective of meaningful contribution, a total of 21 attributes were suggested to be included in the curation of appraisal going forward, which should feed subsequent retraining of the model to adapt to the policy as the business model evolved. From this qualitative search process, 8 of the 21 attributes identified were deemed sufficiently objective to parametrize and model, using the development framework.

*Sentiment model*

The sentiment classifier estimates what the sentiment of a message is, i.e., is it positive or negative or neutral? Messages were annotated manually and validated with the moderator team and founder, with positive examples of messages that fit the label, such as "Hey that was a great game!" expressing excitement and happiness about the game, and negative examples, such as "Man that sucked" expressing disappointment or sadness about something being examples of messages that fit a negative label. Neutral sentiment would be used to annotate messages with no emotional or opinionated content, such as "Yeah that's just the way it is." These serve as a reference for annotators to draw a distinction.

## Results

*Qualitative context*

The community is still young and is characterized by the founders' and original character value sets, which the community moderators described as inclusive, welcoming, and relationship-driven. There were 11 working groups in the community based on allocated roles, of which there are more than 30. From the semi-structured interviews, this culture could be described by the following traits from cultural science:

The people consist of the identified persona groups cryptos, game fans, and casuals, alongside the moderators and the Reality+ developer and founder team. The community language was generally very mature and unique to the specific game and crypto, covering the breadth of the community. The collaborative practices were informally implemented, not consistently managed, and only sometimes documented to evidence consistent coverage. The moderator team and founder appeared to have a very good understanding of the risk to the franchise if onboarding, contribution management, learning, and moderation were not further developed and institutionalized as the franchise prepared for a hypergrowth phase.

The infrastructure technologies consisted of devices (currently PCs only), blockchain, and metaverse technologies relevant to each metaverse chosen by Reality+ and the associated metaverse AI and XR functionality present (Park & Kim, 2022). The institutional technologies identified to underpin the culture were (1) NFTs minted, sold, or awarded (airdropped) by Reality+ for meaningful contributions; (2) Avatars, many of which could only be earned through community contribution; (3) Emotes (expression for doing a good job – emotional expression and visual emoji); (4) Pandaks – an in-game virtual currency used as in-game currency utility token as a free-to-play currency pegged to US$, only accessible through contribution earning in-game; (5) Founders token (FT), which could be bought, but also earned, with limited circulation.

We tested if we could use the rapid model framework to guide us in understanding the main cultural drivers creating meaningfulness in this community. However, the chats were too ambigous to deliver acceptable results, revealing an unimpressive Krippendorff alpha (Krippendorff, 2018) of 0.254 among a panel of 4 humans per step 10 in the development framework.

Several symbolic cultural elements were identified in the community interview process: (a) Due to Doctor Who's rich history, storytelling emerged as a pivotal aspect of community significance. (b) Knowledge and education varied across individuals, with potential enhancements via interactive tutorials and persona-specific onboarding. (c) While there was an active push for innovation, it was sometimes limited by brand intellectual property guidelines. Yet, the community's culture remained unaffected, possibly due to a reward system emphasizing contribution efforts over results. (d) Community behaviors and contributions were primarily influenced by the intrinsic characteristics of the Doctor Who character, and a contribution and reward system centered around NFTs and perks. This system rewarded community growth activities and promoted kindness, inclusivity, and fun. Minimal negative behavior was noted, possibly attributed to robust moderation and the community's manageable size.

Using an archetype model (Cameron & Quinn, 2011), the "Fan" personas had a strong sense of belonging, loyalty, and emotional attachment, reminiscent of a Clan culture. They formed communities where shared passion, narratives, and experiences were central, similar to "strategizers" in the typology presented by (Kahila et al., 2023).

The other key persona group, "Crypto" was more aligned with a Market culture. Their main motivation was often profit, and they operated in a competitive environment where information, timing, and strategic positioning were crucial for success, aligning with the perspectives of (Egliston & Carter, 2023) about the characteristics of cryptogaming. This persona group differs from the typology presented by (Kahila et al., 2023), yet appears to be relevant in blockchain-based games, where DeFi and token economics are an integrated part of the game experience. Given the NFT market's innovative and nascent nature, there might also be Adhocracy elements for both persona types, requiring adaptability and willingness to take risks. The third persona group "Casuals" align with Kahila et al.'s typology "Casual metagamers".

*The model development framework*

The complexity of prompt engineering was demonstrated in the use case, where several iterations were needed for the more difficult classification challenges in moderation and contribution.

Intent model

|  | Precision | Recall | F1-score | N messages |
|---|---|---|---|---|
| Crypto | 0.92 | 0.80 | 0.86 | 41 |
| Fan | 0.93 | 1.00 | 0.96 | 40 |
| Casual | 0.86 | 0.91 | 0.89 | 35 |
| Accuracy |  |  | 0.91 | 116 |
| Macro avg across | 0.90 | 0.91 | 0.90 | 116 |

| Weighted avg | 0.91 | 0.91 | 0.90 | 116 |
| --- | --- | --- | --- | --- |
| **Table 1. Intent model results** | | | | |

The model performance results were production-ready after three iterations of fine-tuning with an estimated model development time of 18 hours over two weeks to allow for processing time for the manual labeling of the 463 chat messages. After fine-tuning the intent model, we ran it on the full database initially supplied by the moderator team. Excluding all empty messages and those from well-known bot accounts, for a total of 59,910 messages across 1,121 active users in 10 channels, the messages were tokenized per OpenAI's API guidance. The full loop took 6 hours to complete, and results showed that approx. 52% of the entire Discord conversation was casual chatter, 25% was related to the gaming universe, and approx. 18% was related to crypto. Categorizing users with three or more messages classified as "crypto" as Crypto Enthusiast personas, users with three or more messages classified as "fan" as Fan personas, and users who are neither Crypto Enthusiasts nor fans as Casual personas suggested 343 unique (pseudonymous) IDs, or app 31% of the active community were Crypto Enthusiasts, 243 or 22% Fans and 716 or 64% Casuals.

*Moderation model*

|  | Precision | Recall | F1-score | N messages |
| --- | --- | --- | --- | --- |
| Toxic | 0.95 | 0.99 | 0.97 | 106 |
| Spam | 1.00 | 0.89 | 0.94 | 9 |
| Not_toxic_not_spam | 0.99 | 0.98 | 0.99 | 268 |
| Accuracy |  |  | 0.98 | 383 |
| Macro avg | 0.98 | 0.95 | 0.97 | 383 |
| Weighted avg | 0.98 | 0.98 | 0.98 | 383 |
| **Table 2. Moderation model results** | | | | |

Given the use case of this moderation tool, we wanted to minimize Type II (false negative) errors, as these are by far the most damaging to culture. Suppose the model wrongly classifies a non-toxic message as toxic. In that case, it is a minor annoyance to the moderator with the additional curation effort; in this case, less than 2 out of 100 flags, while wrongly classifying a toxic message as non-toxic, could be very detrimental to users and drive them away from the community. As the model is implemented, the model should be retrained regularly to further reduce the false negative flags.

*Contribution model*

|  | precision | recall | f1-score | support |
|---|---|---|---|---|
| na | 0.89 | 0.93 | 0.91 | 156 |
| onboarding | 0.75 | 0.9 | 0.82 | 10 |
| knowledge_tcg | 0.57 | 0.5 | 0.53 | 16 |
| knowledge_fan | 0.67 | 0.6 | 0.63 | 10 |
| knowledge_crypto | 0.5 | 0.25 | 0.33 | 4 |
| content | 0.71 | 0.71 | 0.71 | 7 |
| moderation | 0 | 0 | 0 | 1 |
| suggestion | 0.5 | 0.29 | 0.36 | 7 |
| accuracy |  |  | 0.83 | 211 |
| macro avg | 0.57 | 0.52 | 0.54 | 211 |
| weighted avg | 0.82 | 0.83 | 0.82 | 211 |
| **Table 3. Moderation model results** | | | | |

The initial contribution model proved challenging to develop. We first used a zero-shot prompt with the text-DaVinci-003 model to classify messages, but manual checks revealed inadequate quality, potentially due to a vague prompt and random message sampling. Building on a pre-established contribution taxonomy, our second strategy yielded only some satisfactory results. Eventually, recognizing that chat contributions often occur in dialogues, we incorporated the context of the two preceding messages into the model, considerably improving accuracy. While the results table demonstrates the model's ability to detect community contributions, there is room for improvement. Amplifying the training data could enhance outcomes, but the current model was deemed adequate for production, aiding in moderating and annotating contributions through future curation.

*Sentiment model*

| The initial model for benchmarking | Precision | Recall | F1-score | N messages |
|---|---|---|---|---|

x

| | | | | |
|---|---|---|---|---|
| delivered acceptable performance out of the box, roughly equal to the human baseline for sentiment agreement levels (Lexalytics, 2023; Ribeiro et al., 2016). | | | | |
| Positive | 0.75 | 0.75 | 0.97 75 | 32 |
| Neutral | 0.69 | 0.71 | 0.70 | 28 |
| Negative | 1.00 | 0.93 | 0.97 | 15 |
| Accuracy | | | 0.77 | 75 |
| Macro avg | 0.81 | 0.80 | 0.81 | 75 |
| Weighted avg | 0.78 | 0.77 | 0.78 | 75 |
| **Table 4. Sentiment model results** | | | | |

*Replication of model framework to Reality+ ambit*

In advancing Reality+'s vision for a universally harmonized transmedia platform with a consistent ethical goal for metaverse use, we tested our framework on another game within the Reality+ portfolio: Thunderbirds.

Operated by ITV in the Sandbox metaverse, Thunderbirds has its distinct contribution mechanism inherited from that platform. Testing with 102,000 Thunderbird Discord chats revealed similar results for intention, sentiment, and moderation models as with DWWA.

However, the contribution model lagged. Unlike DWWA, where series of messages were analyzed for context, the same sequential approach faltered for Thunderbirds. Discussions with the Thunderbird moderators attributed this to the community's less defined contribution patterns and cultural nuances. Yet, given the transactional nature of the Sandbox Metaverse, the model required significant adjustments. With the Thunderbird community already acquainted with SAND's game rewards, a pivot to community-focused contributions was deemed to be vital, however the overarching approach was considered apt for refining the model with more curated data, so it was decided to implement this model and retrain it regularly based on data from the contributions going forward.

*Impact on agency cost per wallet*

Current costs without the new system per Wallet; The community team managed 250,000 wallets without the system with the app 30 moderators that worked part-time. From the founder, we understood the cost of moderation could be set to USD 50 per hour, and moderators worked half-time. Current Cost per Wallet per Day = Total Current Daily Agency Costs. Current Cost per Day were estimated as 30 x 8 x 50 / 2 = USD 6,000 in total agency cost per day, equal to USD 0,024 per wallet.

The new system incurs a daily cost of USD 5 for regular use, with the highest observed cost being USD 8 for analyzing 60,000 chats during the development, which took 300 hours, translating to USD 15,000 or an amortized daily cost of USD 41.10 over a year. Consequently, the total daily system expenditure is USD 66.10. With this system, the moderation team is able to efficiently manage 5,000,000 wallets at 0.00001322 USD each, marking a cost reduction of 1,815 times. Without the system, handling such a wallet volume would cost USD 120,000 daily. Hence, the system dynamics imply a 95% cost reduction on adoption.

## Discussion

NFTs sparked a revolution in digital ownership, and blockchain-based business models, including DAOs, have found a product-market fit in the cultural economy, reaching from collectors to auctions to media and entertainment, including the metaverse (Messari, 2023).

The cultural economy is characterized by pronounced demand unpredictability, and the quest for tools to mitigate risk in decision-making is paramount (Towse & Hernández, 2020). As our experiment shows, AI can reduce agency costs. Still, there is also a risk of increased agency risk (Sidorova & Rafiee, 2019), but DLT/blockchain can forge a framework suitable for the governance and handling of the critical data that AI systems gather, preserve, and employ in this creator economy.

Our experiment successfully demonstrated the feasibility of implementing GPT-based models to sustain cultural production as an optimized internal control to improve moderation effort significantly as a community scales.

Combining blockchain's affordances with generative AI for cultural production appears to be a promising and powerful mix, which can also make reward practices more focused. While we have not yet seen the results of fully implementing the balanced contribution model in blockchain-based gaming, our findings suggest that the enforcement of a "proof of contribution" model can form a distinguishable and balanced subpopulation of socially networked non-kin with common "institutions of language" and rules to produce meaningfulness within a contextualized niche, which is strong enough to enable large-scale expansion, and, at the same time, avoid some unintended consequences of gaming.

GPT presents a step change to model development, where, based on processing time and model performance, the technology allowed us to create strong models in weeks that would otherwise have taken months with many resources in a traditional programming language setting.

Previous analysis of the Reality+ NFT marketplace revealed no material wash trading, so assuming one unique Discord ID represents one of the wallets connected to this NFT-based game, of which only 1,121 or app 1,4% access the Discord community, only 686, or less than one pct of the total number of wallets are relevant from a community-building and commercial perspective. Compared to financial data from Reality+, these results proved very realistic.

The results of the moderation modeling are quite promising and suggest that automated (mass) content moderation is feasible within the planning horizon of Reality+. However, even though the agency cost is reduced by 95 pct in this startup case, the results from our use case are not mass volume. It remains to be tested whether a cost-efficient model can be implemented in very large communities or whether it becomes prohibitively expensive to use a closed-source, open-access API service such as OpenAI.

Likewise, we opted to test OpenAI's GPT suite as a "quick win" to introduce advanced analytics to a young gaming community. Any of the LLMs in OpenAI's GPT 3.5 suite are likely much too advanced for the analytical challenges presented here, as the challenges are traditional classification problems, where less complex models perform. Yet, the project demonstrated to the moderator team how analytics could help them manage the community cultures during a hypergrowth phase. Evaluation feedback suggests that many other use cases will be moved forward, including incorporation of memory to the LLM through indexing and vectorizing the full histories behind the gaming characters to allow for faster and better support and knowledge management, as recently described in grey literature (IBM, 2023).

The contribution model findings underscore the potential of analytics in fostering and preserving shared cultural and ethical values across a broader ecosystem of interrelated communities, yet with some implementation effort, as the Thunderbird model endeavor suggested.

In our use case, a level of centralized control is required to manage customer brand IP. Yet, the mix of DLT and AI demonstrates the potential to implement a decentralized creator economy curated with intersubjectivity to reduce the risk of gaming a purely objective social algorithm.

## Conclusion

By driving cooperation and coordination, traditional firms can indirectly enhance performance by creating a strong culture (Murphy et al., 2013). In this context, culture is not something you are but something you do as a network of live interactions striving towards a common objective. We can recognize the effects of a strong culture in successful enterprises and when it is toxic or missing.

Yet, increasingly, as we have seen in this case, large loosely tied communities develop online, where individuals work without traditional hierarchical restrictions and frequently with little to no exchange of direct economic value. Such networks create successful cultures and scale without top-down organizational steering. It has been suggested (Coyle, 2018), and also showcased in this paper, that such networked communities collaborate through an effective cultural code that exists beyond the control of managers, consisting of three elements: (1) signs of connection produce ties of identification and belonging, (2) mutual risk-taking behaviors foster trusting collaboration, and (3) narratives produce shared values and objectives.

Aligning to current research agendas in Blockchain, Metaverses, and DAOs, this paper explores culture from an evolutionary perspective in semi-decentralized web3-based gaming communities leveraging a novel blockchain-based business model that we call "proof of contribution," designed to reduce unintended consequences of addictive P2E gaming models in metaverses.

Using a case study approach, we asked the following research question (RQ): "*Are LLMs an option in identifying, nurturing, and sustaining culture in gaming communities going through hypergrowth?*"

We hypothesized that (1) pseudonymous actor intent can be predicted based on chat activity, (2) generative AI can make mass content moderation effective, (3) a common moral mission can be implemented effectively with advanced analytics, and (4) there is an unrealized automation potential that can significantly reduce moderation effort and compliance burden in managing community culture as web3 businesses scale. All hypotheses were confirmed, and we answered the RQ by presenting. (1) a pragmatic framework for rapid model development leveraging open-access LLMs to accelerate the use of analytics in an online community, (2) a number of industrial-ready trained model artefacts using OpenAI's API-based GPT-suite for decentralized communities wishing to nurture their culture while scaling in a collaborative web3-based economy/metaverse construct, including (a) an intent classification model to predict pseudonymous community actor intent based on community behavior, (b) a moderation model to signal unwanted behavior for curation, (c) a contribution model to signal proof of desired contribution for further curation and reward based on a common moral mission to mitigate emerging risks in addictive blockchain-based gaming to reduce moderator workload and cost-effectively manage community culture in hypergrowth scenario, and finally (d) a community sentiment model.

While these functionalities are not new, the application in startup governance and the usage to manage group formation and production of culture in pseudonymous, decentralized communities planning for hypergrowth is, to the best of our knowledge, unique.

Key contributions of our framework to the field of gaming analytics and production include (i) deciphering intentions in pseudonymous economies: As communities evolve into new metaverse business models, the capability to identify the intent of pseudonymous actors becomes crucial, informing both economic interactions and cultural evolution; (ii) cultural moderation in digital economies: Our design pinpoints both toxic and desired behaviors. This nuanced understanding facilitates the curation of rewarding mechanisms aligned with a novel blockchain-based gaming contribution principle, ensuring cultural and economic alignment; (iii) sentiment analysis and cultural pulse: The ability to assess community sentiment provides real-time feedback on cultural shifts, an invaluable tool for both economic modeling and cultural preservation; and (iv) facilitating targeted growth with shared values: Our models strengthen the capacity to onboard specific persona groups, which significantly supports rapid growth with a consistent cultural mission, essential for sustainable economic activity in interconnected community ecosystems.

We contribute to the transmedia storytelling discourse with (i) the application of a contribution-based business model focused on collaboration, teamwork, and contribution around a common moral mission when transforming from web2 to web3 and metaverses, (ii) the governance of blockchain-based business models by introducing generative AI-based supported moderation policies, (iii) meta-sustainability by internalizing sustainability values at an individual level while developing a collective mindset that understands the importance of doing the right things the right.

Our findings underline the transformative potential of generative AI in cultural economics. Beyond improving gaming experiences or crafting immersive metaverse games, AI's capability to decode and shape gaming community cultures presents a new frontier in the innovative design of economic systems. Therefore, this research demonstrates the practical use of LLMs in managing digital communities and significantly broadens the scope of their application in cultural economics.

## Limitations and further work

We mainly utilized GPT3.5 for our research instead of the newer GPT4. Our priority was to enhance moderators' capabilities urgently, so we tested an API-based solution without exploring advanced systems like Google's BERT. Ideally, different models should be compared for accuracy on the same dataset, especially as datasets improve and require model retraining. Also, this study did not delve into gaming and metaverse infrastructures; Reality+ is still a startup developing its platform-agnostic metaverse strategy centered around Discord as relates community, and although the models proved highly useful in this stage of their growth journey, it remains to be tested if the models can perform cost-effectively as the volume increases. Additionally, the potential of GPT models in shaping decentralized business and algorithmic governance needs exploration, as and when Reality implements a financial model for contribution and decentralized ownership. Finally, our focus was on unique IDs for avatars in Discord. Still, a single individual might use multiple IDs, and verifying ID independence is crucial as the community grows to prevent system manipulation.

## Acknowledgements

This work is partially funded by a grant provided by the Danish Research and Innovation Council as administered and awarded by Copenhagen FinTech for the project Republic of Reality+. Additionally, the project received in-kind funding by Reality Plus ApS and Doerscircle Pte. Also, this research has received funding from the European Union's Horizon 2020 research and innovation programme, within the OpenInnoTrain project under the Marie Sklodowska-Curie grant agreement n°823971.